\documentclass[submission, Phys]{SciPost}

\hypersetup{
    colorlinks,
    linkcolor={red!50!black},
    citecolor={blue!50!black},
    urlcolor={blue!80!black}
}

\usepackage[bitstream-charter]{mathdesign}
\DeclareSymbolFont{usualmathcal}{OMS}{cmsy}{m}{n}
\DeclareSymbolFontAlphabet{\mathcal}{usualmathcal}

\begin{document}

\begin{center}{\Large \textbf{
Electric and Magnetic Tau Dipole Moments Revisited\\
}}\end{center}

\begin{center}
Gabriel A. Gonz\'alez-Sprinberg\textsuperscript{1},
\end{center}

\begin{center}
{\bf 1} Faculty of Sciences, Universidad de la Rep\'ublica, Uruguay
\\
* Gabriel.Gonzalez-Sprinberg@fcien.edu.uy
\end{center}

\begin{center}
\today
\end{center}


\definecolor{palegray}{gray}{0.95}
\begin{center}
\colorbox{palegray}{
  \begin{tabular}{rr}
  \begin{minipage}{0.1\textwidth}
    \includegraphics[width=30mm]{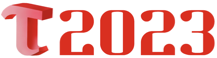}
  \end{minipage}
  &
  \begin{minipage}{0.81\textwidth}
    \begin{center}
    {\it The 17th International Workshop on Tau Lepton Physics}\\
    {\it Louisville, USA, 4-8 December 2023} \\
    \doi{10.21468/SciPostPhysProc.?}\\
    \end{center}
  \end{minipage}
\end{tabular}
}
\end{center}

\section*{Abstract}
{\bf
Precise measurements of magnetic and electric dipole moments are important tests of the 
Standard Model and beyond Standard Model physics, particularly for the electron and the
 muon. However, the situation presents distinctive 
 challenges when dealing with the tau lepton  due to its very short lifetime
 and relatively high mass. Here, we 
 review the theoretical predictions and  experimental measurements of both the 
 anomalous magnetic and electric dipole moments of the tau lepton.
}

\vspace{10pt}
\noindent\rule{\textwidth}{1pt}
\tableofcontents\thispagestyle{fancy}
\noindent\rule{\textwidth}{1pt}
\vspace{10pt}

\section{Introduction: Tau Dipole Moments}
\label{sec:intro}
Tau dipole moments (DM) have been investigated since the discovery of this high-mass
lepton \cite{1975_Perl_PRL_35}.
It was speculated that the DMs could play a role similar to those the 
electron and the muon DMs 
were playing in the Standard Model (SM) and beyond the SM  physics 
(BSM). However, the measurements for both the magnetic and electric DMs 
of the tau slowly evolved, and the measurements are still very far 
from the SM predictions, in particular  for the former \cite{pdg}.
In contrast, the electron magnetic moment is one of the highest 
precision measurements in physics and the theory for this DM provides a strong test for QED. 
Additionally, the 
result reported by J. Schwinger \cite{Schwinger:1948iu} for the electron
laid the foundations for quantum field theory as a tool 
for understanding nature at small scales. 
We  begin presenting the electric 
 DM in the next section, followed by the magnetic DM, and, finally, some conclusions in the current 
 experimental and theoretical prospects.

\section{Tau Electric Dipole Moment}
Electric dipole moments (EDM) were introduced as a probe of parity and 
time reversal invariance in particle physics
\cite{LANDAU1957127}. In quantum field theory, 
time reversal violation also implies CP violation. 
For a lepton, the EDM is the first-order CP-odd interaction with the photon field:
\begin{equation}
H_{EDM}= \frac{i}{2}d\,\bar{l}\gamma^5 \sigma^{\mu\nu}l\, F_{\mu\nu},
\end{equation}
where $l$ is the lepton field, and $F_{\mu\nu}$ is the electromagnetic field tensor.
This equation, in the non-relativistic limit, gives the usual first-order interaction with 
an electric field: 
\begin{equation}
H_{EDM}=\,-\,\vec{d} \,\cdot\, \vec{E}.
\end{equation}
For the tau, 
the high   $m_\tau = (1776.86 \pm0.12) \,MeV$ may be advantageous because the EDM tensor structure flips 
chirality and then it is proportional to the chirality flipping particle's mass.  Moreover, many BSM predictions 
for the  tau EDM are proportional to some power $n$ of the particle mass over the scale of new physics,
{\it i.e. }  $(m_\tau/m_\Lambda)^n$. When compared to the electron and muon, being that
$(m_\tau/m_e)^2 \simeq 10^7$ and  $(m_\tau/m_\mu)^2 \simeq 300$, tau physics looks like a promising  
scenario for searching for EDM signals from BSM physics. On the other hand, the very short lifetime 
 $(2.903 \pm0.005)  
\times 10^{-13}\, s$ poses an experimental challenge, necessitating a
completely different approach to searching for the tau EDM 
should be followed.  Recently, stringent bounds on the electron and muon EDM were obtained in  
experiments \cite{pdg} from CP-odd
observables. Besides, many BSM models were ruled out or their space parameter was strongly limited
when confronted with the experiments. 
The PDG bounds \cite{pdg} for the leptons EDM are:
\begin{equation}
| d_{e} | < 1.1 \times 10^{-29} \,\,  \text{\it{e cm} at CL=90.0\%}	
\,\,,\,\,\,
| d_{\mu} | < 1.8 \times 10^{-19}\,\,  \text{\it{e cm} at CL=95\%},
 \end{equation}
 \begin{equation}
 -1.85 \times 10^{-17} \,\, \it{e cm} < Re(d_{\tau}) < 6.1 \times 10^{-18} \,\, \it{e cm}, 
\label{red} \end{equation}
 \begin{equation}
-1.03 \times 10^{-17} \,\, \it{e cm} < Im(d_{\tau}) <2.30 \times 10^{-18} \,\, \it{e cm}.\label{imd}
\end{equation}
Tau bounds are much weaker, particularly when compared to the electron one.
 Limits of the tau EDM slowly evolved over the past decades, with the present bounds \eqref{red} and
 \eqref{imd} derived from the work of K.Inami {\it etal} at BELLE \cite{Belle:2021ybo}.
 Additionally, due to the tiny value of the EDM for leptons 
 in the SM, arising from three-loop diagrams \cite{Shabalin:1978rs},
 any non-zero signal should  
 indicate new physics. The EDM depends on the underlying mechanisms of CP-violation of the BSM 
 physics and can be obtained by a one-loop vertex correction in these models \cite{POSPELOV2005119}. 
 In the effective lagrangian approach,  the EDM is represented 
 by a dimension six operator \cite{Buchmuller:1985jz}.
 By naively scaling the EDM by mass, and considering 
the SM computation for the electron EDM, for example, one might expect
$d_{\tau}=\frac{m_\tau}{m_e}d_{e}$, where the latter is expected to be on the 
order of $10^{-38}$ \cite{POSPELOV2005119}. Note that this value is 21 
orders of magnitude below the current experimental bounds (see \eqref{red} and
 \eqref{imd}).

\noindent Note that both CP-odd and CP-even observables may receive an EDM contribution. However, 
for CP-even observables, contributions are proportional to $|d_\tau|^2$. This 
is the case for the contributions coming from  the leading diagrams 
for these kind of observables,  shown in Fig. 1 for some processes. These correspond to 
cross sections, angular distributions, and decay widths, shown in this figure 
in  
(a) $\sigma(e^+ e^- \rightarrow \tau^+ \tau^-)$, 
(b) $\sigma(e^+ e^- \rightarrow \tau^+ \tau^- e^+  e^-)$,
(c) $\sigma(e^+e^- \rightarrow \tau^+ \tau^- \gamma)$, 
(d) $\Gamma (Z \rightarrow \tau^+ \tau^- \gamma)$,
(e) $\sigma(\gamma\gamma \rightarrow \tau^+ \tau^-)$, 
(f)  $\Gamma (\tau \rightarrow l^+ \nu_\tau  \bar{\nu}_l  \gamma)$ 
for example.
\begin{figure}[htbp]
\centering
\includegraphics[width=0.8\textwidth]{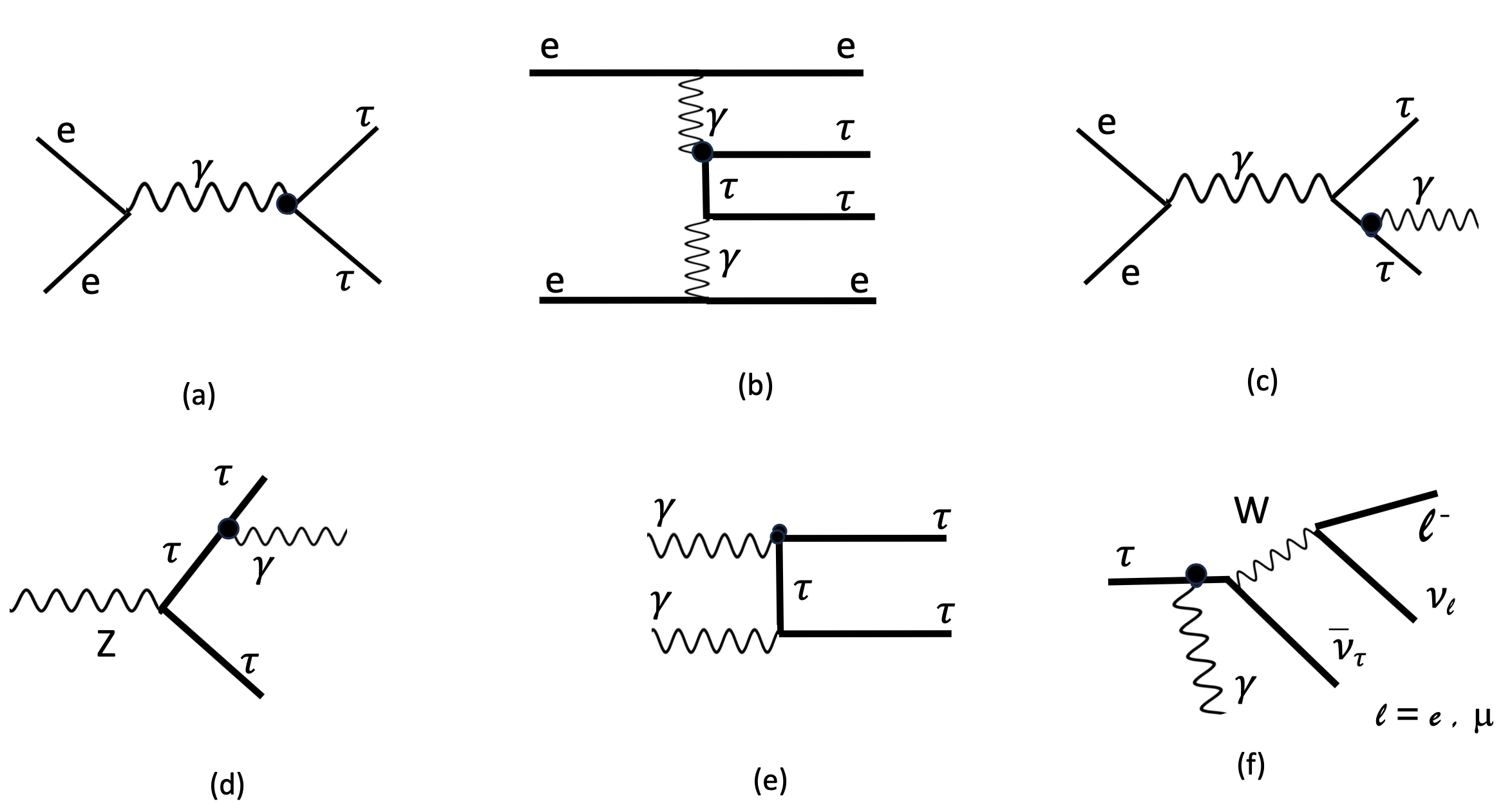}
\caption{Some processes where the $\tau$ dipole moments  may contribute.}
\label{ref}
\end{figure}
\noindent These were 
studied or proposed for experiments at PETRA, LEP, LEP2, BABAR, BELLE, BELLE2, and the
LHC (see, for example, \cite{Bernreuther:2021elu, Haisch:2023upo,Fomin:2018ybj}).
 On the other hand, CP-odd observables proportional to the tau-EDM can be studied
in tau-pair production by a s-channel photon diagram. These can be analyzed in the
spin terms (spin linear or spin-spin correlation terms) that modify the 
angular distributions of the decay products, or in expectation values of 
triple spin-momentum observables. The latter were used by K.Imani at BELLE 
following the ideas in
\cite{Bernreuther:2021elu}. 
It is expected that the high statistics from B-factories will allow for 
better bounds on the tau EDM, potentially reaching levels where some BSM physics might be detectable.
\section{Tau Magnetic Dipole Moment}
\noindent The first-order leading contribution in QED \cite{Schwinger:1948iu} for the 
electron magnetic moment anomaly, 
leads to the flavor-independent expression:
\begin{equation}
a = \frac{g - 2}{2} = \frac{\alpha}{2 \pi}\simeq 0.00116.
\end{equation}
This computation was updated in \cite{Eidelman:2007sb} for the tau,  
including weak and hadronic contributions:
\begin{equation}
a_{\tau} = 117721(5) \times 10^{-8}.
\end{equation}
\noindent The PDG bound for the tau anomaly is
$ -0.052 < a_{\tau} < 0.013 \,\, (95\% \, CL) $. This value, taken from \cite{DELPHI:2003nah} 
more than two decades ago, was obtained from total cross-section
data, such as the one in fig. 1 (a) for DELPHI. 
Other processes, as shown in fig. 1,
were also considered by other authors to 
set bounds on the tau anomaly. In \cite{Gonzalez-Sprinberg:2000lzf}, a global 
and model-independent analysis through effective lagrangians of SLD and LEP 
data shows that a stringent bound for 
BSM contributions to the tau magnetic anomaly can be obtained:
$- 0.007 < a_{\tau}^{BSM} < 0.005 \,(2\sigma)$. The case for the tau anomaly 
is very different from that of the electron or muon. While the SM theoretical 
prediction is also well-known, the experimental precision for the tau anomaly is still 
lacking, and even the sign is unknown.
The tau promptly decays, and it is through the angular distribution 
of the secondary particles of the decay that the anomaly can be traced back. 
Generally, total cross-section may provide an indirect bound on the 
anomaly. Moreover, spin or spin-spin correlations terms in the final tau pairs could serve 
as sensitive observables for detecting the tau anomaly. 
Several of the studied observables do not have all three particles
on-shell. This means that what is actually being measured or bound is not the 
SM prediction, where both $\tau$ and photon are on-shell, but a bound on possible
BSM contributions to the form factor of this off-shell vertex. Furthermore, it is 
usually assumed that not all the form factors that parameterize this off-shell 
vertex are considered. More recently, proposals 
for many colliders, particularly for B-factories experiments have been put forward.
For example, in \cite{Bernabeu:2007rr,Bernabeu:2008ii},  it was shown that  
the QED magnetic form factor for $q^2=M_{\Upsilon}^2$ can be obtained from asymmetries 
constructed with longitudinal and transverse polarizations of the produced taus for polarized beams.
All these studies have  been considered in \cite{Crivellin:2021spu}, which shows that feasible 
BSM values of the tau anomaly at the level $10^{-6}$ could be probed with a polarization upgrade 
of SuperKEKB.

\section{Conclusion}

Tau DMs remain an attractive area of research, with potential new results expected in the near future.
Recent experiments related to 
the electric DM have shown improvement due to the high-statistics data 
from B-factories. For the tau magnetic anomaly, the experimental limits are 
still orders of magnitude above 
the SM prediction, 
and even the sign of the tau anomaly remains unknown. The magnetic DM experiments 
are making  progress, and new experimental ideas are emerging. Accessing the 
$q^{2}=0$ SM prediction is challenging due to the very short lifetime of the tau-lepton.
However, for  $q^{2} \simeq M_{\Upsilon}^{2}$ at B-factories, 
 some promising ideas may help to establish bounds on the magnetic form 
 factor at B-factories at levels where BSM effects could appear.  
As Martin Perl \cite{arXiv:9812400} stated many decades ago in 
{\it Dreams and odd ideas in tau research}: "... It would be very nice to measure 
$\mu_\tau$ with enough precision to check this ({\it the Schwinger term $\alpha /2 \pi$}), 
as it was checked for the $e$ and the $\mu$ years ago. At present such precision is a dream.”
The tau DMs may still yield impressive results due to the BSM physics sensitivity 
to this high-mass lepton and to the high statistics available from colliders experiments.

\section*{Acknowledgements}
J.Vidal, J. Bernabeu, and A. Santamaria are acknowledged for illuminating discussions 
over the last years. PEDECIBA-Uruguay support is also gratefully acknowledged.

\nolinenumbers


\begin{thebibliography}{10}
\providecommand{\url}[1]{\texttt{#1}}
\providecommand{\urlprefix}{URL }
\expandafter\ifx\csname urlstyle\endcsname\relax
  \providecommand{\doi}[1]{doi:\discretionary{}{}{}#1}\else
  \providecommand{\doi}{doi:\discretionary{}{}{}\begingroup \urlstyle{rm}\Url}\fi
\providecommand{\eprint}[2][]{\url{#2}}

\bibitem{1975_Perl_PRL_35}
M.~L. Perl \emph{et~al.},
\newblock \emph{Evidence for anomalous lepton production in e+ - e- annihilation},
\newblock Phys. Rev. Lett. \textbf{35}, 1489 (1975),
\newblock \doi{10.1103/PhysRevLett.35.1489}.

\bibitem{pdg}
R.~L. Workman and Others,
\newblock \emph{{Review of Particle Physics}},
\newblock PTEP \textbf{2022}, 083C01 (2022),
\newblock \doi{10.1093/ptep/ptac097}.

\bibitem{Schwinger:1948iu}
J.~S. Schwinger,
\newblock \emph{{On Quantum electrodynamics and the magnetic moment of the electron}},
\newblock Phys. Rev. \textbf{73}, 416 (1948),
\newblock \doi{10.1103/PhysRev.73.416}.

\bibitem{LANDAU1957127}
L.~Landau,
\newblock \emph{On the conservation laws for weak interactions},
\newblock Nuclear Physics \textbf{3}(1), 127 (1957),
\newblock \doi{https://doi.org/10.1016/0029-5582(57)90061-5}.

\bibitem{Belle:2021ybo}
K.~Inami \emph{et~al.},
\newblock \emph{{An improved search for the electric dipole moment of the $\tau$ lepton}},
\newblock JHEP \textbf{04}, 110 (2022),
\newblock \doi{10.1007/JHEP04(2022)110}.

\bibitem{Shabalin:1978rs}
E.~P. Shabalin,
\newblock \emph{{Electric Dipole Moment of Quark in a Gauge Theory with Left-Handed Currents}},
\newblock Sov. J. Nucl. Phys. \textbf{28}, 75 (1978).

\bibitem{POSPELOV2005119}
M.~Pospelov and A.~Ritz,
\newblock \emph{Electric dipole moments as probes of new physics},
\newblock Annals of Physics \textbf{318}(1), 119 (2005),
\newblock \doi{https://doi.org/10.1016/j.aop.2005.04.002}.

\bibitem{Buchmuller:1985jz}
W.~Buchmuller and D.~Wyler,
\newblock \emph{{Effective Lagrangian Analysis of New Interactions and Flavor Conservation}},
\newblock Nucl. Phys. B \textbf{268}, 621 (1986),
\newblock \doi{10.1016/0550-3213(86)90262-2}.

\bibitem{Bernreuther:2021elu}
W.~Bernreuther, L.~Chen and O.~Nachtmann,
\newblock \emph{{Electric dipole moment of the tau lepton revisited}},
\newblock Phys. Rev. D \textbf{103}(9), 096011 (2021),
\newblock \doi{10.1103/PhysRevD.103.096011}.

\bibitem{Haisch:2023upo}
U.~Haisch, L.~Schnell and J.~Weiss,
\newblock \emph{{LHC tau-pair production constraints on $a_\tau$ and $d_\tau$}},
\newblock SciPost Phys. \textbf{16}(2), 048 (2024),
\newblock \doi{10.21468/SciPostPhys.16.2.048}.

\bibitem{Fomin:2018ybj}
A.~S. Fomin, A.~Y. Korchin, A.~Stocchi, S.~Barsuk and P.~Robbe,
\newblock \emph{{Feasibility of $\tau$ -lepton electromagnetic dipole moments measurement using bent crystal at the LHC}},
\newblock JHEP \textbf{03}, 156 (2019),
\newblock \doi{10.1007/JHEP03(2019)156}.

\bibitem{Eidelman:2007sb}
S.~Eidelman and M.~Passera,
\newblock \emph{{Theory of the tau lepton anomalous magnetic moment}},
\newblock Mod. Phys. Lett. A \textbf{22}, 159 (2007),
\newblock \doi{10.1142/S0217732307022694}.

\bibitem{DELPHI:2003nah}
J.~Abdallah \emph{et~al.},
\newblock \emph{{Study of tau-pair production in photon-photon collisions at LEP and limits on the anomalous electromagnetic moments of the tau lepton}},
\newblock Eur. Phys. J. C \textbf{35}, 159 (2004),
\newblock \doi{10.1140/epjc/s2004-01852-y}.

\bibitem{Gonzalez-Sprinberg:2000lzf}
G.~A. Gonzalez-Sprinberg, A.~Santamaria and J.~Vidal,
\newblock \emph{{Model independent bounds on the tau lepton electromagnetic and weak magnetic moments}},
\newblock Nucl. Phys. B \textbf{582}, 3 (2000),
\newblock \doi{10.1016/S0550-3213(00)00275-3}.

\bibitem{Bernabeu:2007rr}
J.~Bernabeu, G.~A. Gonzalez-Sprinberg, J.~Papavassiliou and J.~Vidal,
\newblock \emph{{Tau anomalous magnetic moment form-factor at super B/flavor factories}},
\newblock Nucl. Phys. B \textbf{790}, 160 (2008),
\newblock \doi{10.1016/j.nuclphysb.2007.09.001}.

\bibitem{Bernabeu:2008ii}
J.~Bernabeu, G.~A. Gonzalez-Sprinberg and J.~Vidal,
\newblock \emph{{Tau spin correlations and the anomalous magnetic moment}},
\newblock JHEP \textbf{01}, 062 (2009),
\newblock \doi{10.1088/1126-6708/2009/01/062}.

\bibitem{Crivellin:2021spu}
A.~Crivellin, M.~Hoferichter and J.~M. Roney,
\newblock \emph{{Toward testing the magnetic moment of the tau at one part per million}},
\newblock Phys. Rev. D \textbf{106}(9), 093007 (2022),
\newblock \doi{10.1103/PhysRevD.106.093007}.

\bibitem{arXiv:9812400}
M.~L. Perl,
\newblock \emph{The tau lepton and the search for new elementary particle physics, 5th international wein symposium: A conference on physics beyond the standard model (wein 98)},
\newblock \doi{10.48550/arXiv:9812400}.

\end{thebibliography}
\end{document}